\documentclass[twoside]{amsart}
\numberwithin{equation}{section}
\usepackage{amsmath}
\usepackage{amssymb,amsfonts,amsthm,latexsym,psfrag}
\usepackage{mathrsfs}
\usepackage[english]{babel}

\theoremstyle{plain}
        \newtheorem{theorem}{Theorem}[section]
        \newtheorem{lemma}[theorem]{Lemma}
        \newtheorem{proposition}[theorem]{Proposition}
        \newtheorem{corollary}[theorem]{Corollary}
               
\theoremstyle{definition}

\newcommand {\qkos} {\mbox{${\partial_\nu}$}}
\newcommand {\qce} {\mbox{$\delta_\nu$}}
\newcommand {\qhop} {\mbox{$h_\nu$}}
\newcommand {\qdz} {\mbox{$d_\nu$}}
\newcommand {\qbrst} {\mbox{$\mathscr D_\nu$}} 
\newcommand {\q} {\mbox{\boldmath $q$}}
\newcommand {\p} {\mbox{\boldmath $p$}}
\newcommand {\id} {\operatorname{id}}
\newcommand {\ad} {\operatorname{ad}}
\newcommand {\res} {\operatorname{res}}
\newcommand {\prol} {\operatorname{prol}}

\begin{document}

\title[Singular reduction in deformation quantization]{A homological approach to 
       singular reduction in deformation quantization}

\author{Martin Bordemann, Hans-Christian Herbig and Markus Pflaum}
\address{Laberatoire de Math\'ematique de Haute Alsace, Mulhouse, France}
\email{Martin.Bordemann@uha.fr}
\address{Fachbereich Mathematik, Goethe-Universit\"at, Frankfurt/Main, 
         Germany}
\email{herbig@math.uni-frankfurt.de}    
\address{Fachbereich Mathematik, Goethe-Universit\"at, Frankfurt/Main, 
         Germany}   
\email{pflaum@math.uni-frankfurt.de} 

\dedicatory{Dedicated to Jean-Paul Brasselet on the occasion of his 60th 
        birthday}

\begin{abstract}
We use the method of homological quantum reduction 
to construct a deformation quantization on singular symplectic quotients 
in the situation, where the coefficients of the  moment 
map define a complete intersection. Several examples are discussed, among others one
where the singularity type is worse than an orbifold singularity.
\end{abstract}

\maketitle
\tableofcontents
\section{Introduction}

In hamiltonian mechanics, reducing the number of degrees of freedom of a hamiltonian 
system by exploiting its symmetry is a standard method to determine the dynamics of the
system. Within the language of symplectic geometry, regular reduction has 
been introduced independently by Meyer and Marsden/Weinstein  and is usually called 
Marsden--Weinstein reduction. In \cite{Fed98} and, subsequently, in \cite{BHW} it was 
shown that Marsden-Weinstein reduction has an analog in deformation 
quantization (see \cite{Dito} for an overview on deformation quantization) in case the 
hamiltonian group action satisfies certain  regularity conditions. 
This quantum reduction was used to obtain differentiable star products on 
regular symplectic quotient spaces. The general approach 
followed in \cite{BHW} is known as the BRST-method and goes back to works of 
Batalin, Fradkin and Vilkoviski 
(for an overview and references on classical homological reduction see 
\cite{Stasheff}). 

In the following, we will see that the above method, suitably modified, works also for 
cases of \emph{singular} reduction, where the singular behavior of the moment map is 
``not too bad''. This will yield continuous star products on the corresponding 
singular quotient spaces. Let us be more specific about the premises to be made. 
We will consider a hamiltonian action of a connected and compact Lie group $G$ 
acting on a symplectic manifold $M$ with equivariant moment map 
$J:M\to \mathfrak g^*$, where $\mathfrak g^*$ is the dual space of the Lie algebra 
$\mathfrak g$ of $G$. Let $Z:=J^{-1}(0)$ be the zero set of $J$, it will be also 
called \emph{constraint surface}. Due to the equivariance of $J$, the constraint 
surface is an invariant subset. Let us denote by 
$I(Z)\subset \mathcal C^{\infty}(M)$ the vanishing ideal of $Z$. 
We will assume that the moment map satisfies the following conditions:
\begin{enumerate}
\item the components of $J$ generate $I(Z)$ (generating hypothesis),
\item the Koszul complex on $J$ in the ring $\mathcal C^\infty(M)$ is acyclic 
      (cf.~Section 3).
\end{enumerate}
Substantial work has been done in \cite{AGJ} in order to understand the generating 
hypothesis. Using local normal coordinates for the moment map this issue is reduced 
to a problem in algebraic geometry (cf.~also Section 2). Note that the generating 
hypothesis puts severe restrictions on the geometry of $Z$: it implies that $I(Z)$ is a 
Poisson subalgebra. Using Dirac's terminology we say: $Z$ is \emph{first class}. If the 
Koszul complex is acyclic,  one also says  $J$ is  a \emph{complete intersection} 
(see e.g.~\cite{Bourb10}). Misleadingly, the physicist's denotation is here: 
``$J$ is irreducible''. The question whether a variety is (locally) a complete 
intersection is fundamental in commutative algebra, but there the most interesting 
techniques to determine that rely on the assumption that the base ring is  
\emph{noetherian}, as opposed 
to the ring of smooth functions on a manifold which is the base ring in our considerations. 
So we have to find alternatives and attack this problem directly by providing a simple 
crititerion for $J$ to be a complete intersection (cf.~Theorem \ref{acyckrit}). 
The proof may be interesting in its own right.

If zero is a singular value of the moment map, the constraint surface $Z$ is not a 
smooth manifold, but, according to \cite{SjLerm}, a stratified space. A continuous 
function $f$ on $Z$ is said to be \emph{smooth} if there is a smooth function 
$F\in\mathcal C^{\infty}(M)$ such that $f=F_{|Z}$. The algebra of smooth functions 
$\mathcal C^{\infty}(Z)$ is isomorphic to $\mathcal C^\infty(M)/I(Z)$. It is naturally 
a  Fr\'echet algebra, since it is the quotient of a Fr\'echet algebra by a closed ideal.
In \cite{SjLerm} Sjamaar and Lerman could show that the orbit space of $Z$ under the 
action of $G$ is a stratified symplectic space.  The Poisson algebra  of smooth 
functions on it is naturally isomorphic to the Poisson algebra 
$\mathcal C^\infty(Z)^G/I(Z)^G$. 
Since $Z$ is first class, $\mathcal C^\infty(Z)^{\mathfrak g}$ carries a canonical 
Poisson structure, which is referred to as the \emph{Dirac reduced algebra}. 
Since $G$ is compact and connected, these Poisson algebras are isomorphic. If the 
conditions a) and b) above are true, this Poisson algebra is identified with the 
zeroth cohomology of the classical BRST-algebra (cf.~Section \ref{classred}). 

According to \cite{BHW}, it is relatively easy to find a formal deformation of the 
classical BRST-algebra into a differential graded associative algebra such that the 
cohomology is essentially unchanged (see Section \ref{quantumBRST} and 
\ref{quantumred}), and thus yielding a deformation of the classical reduced Poisson 
structure. In \cite{BHW} some efforts have been made to provide explicit formulas for 
contracting homotopies of the Koszul resolution, which have certain technical 
properties. Using these formulas and techniques from homological perturbation theory,
it was shown that, in the regular case, a differentiable reduced star product can be 
found. Here we use the extension theorem and the division theorem of 
\cite{Schwarzbier} to provide \emph{continuous} contracting homotopies that satisfy 
similar technical assumptions. 

In this way, we obtain the main result of this article. Given a hamiltonian action 
of a compact connected Lie Group on a symplectic manifold such that the moment map 
satisfies conditions a) and b) above, then there exists a \emph{continuous} formal 
deformation of the Dirac reduced algebra, i.e.~a continuous star product on the 
singular reduced space (see Corollary \ref{redprod}). Since it is clear, that a 
situation, where both conditions a) and b) are true, is rather special, we start 
the discussion by giving some examples (cf.~Section \ref{ex}). Needless to say, this will 
show that the theory does not reduce to the regular situation. But, more importantly, there 
are examples where the reduced spaces are not orbifolds, but genuine  stratified symplectic 
spaces. To the authors knowledge, this is the first known instance of such a space admitting 
a deformation quantization. Homological reduction therefore provides a construction method 
for formal deformation quantizations  which works for more 
general singular symplectic spaces than the Fedosov type construction introduced in 
\cite{PflDQSO} for orbifolds.

We have included an appendix providing basic notions of homological perturbation 
theory and two variants of the well known basic perturbation lemma 
(see e.g.~\cite{LambeSt}), which are less universal but fit our purposes. 
The perturbation lemma \ref{BPL1} is also implicit in Fedosov's construction 
\cite{Fed94}. 

Througout this paper we shall use the following conventions. Unless otherwise stated,
all complexes are \emph{cochain} complexes in the category of $\mathbb K$-vector 
spaces, $\mathbb K$ being $\mathbb R$ or $\mathbb C$. The shift $V[j]$ of a graded 
vector space $V=\oplus_i V^i$ is defined by $V[j]^i:=V^{i+j}$. If not said 
otherwise, maps of graded vector spaces are of degree zero. Concerning symplectic 
structure, moment maps, star products etc.~we adopt the conventions of 
\cite{BHW}. The formal parameter $\nu=i\lambda$ stands for $i\hbar$.
\vspace{3mm}

\paragraph{\bf{Acknowledgements.}}  
The authors would like to thank Markus Hunziker for stimulationg discussions and
drawing our attention to important references concerning commuting varieties. We thank
Pawe{\l} Doma\'nski for explaining the notion of a split, 
and Richard Cushman, Marc Gotay, Nolan Wallach and Patrick Erdelt for helpful advice. 
M.P. and H.-C.H. gratefully acknowledge support by Deutsche Forschungsgemeinschaft. 
H.-C.H. also acknowledges a travel stipend by  Hermann Willkomm-Stiftung.
\section{Examples}\label{ex}
Before we start to explain the general machinery let us provide some examples of 
hamiltonian $G$-spaces, which satisfy the generating and the complete intersection 
hypothesis.
In general, it is not at all a trivial matter to check, whether the generating 
hypothesis is true. The following is based on results of the seminal article 
\cite{AGJ}. We begin the discussion with the most simple case, where $G$ is a torus. 

\subsection{Hamiltonian torus actions}
In \cite{AGJ} it was proven that for a moment map $J:M\to \mathfrak g^*$ of a torus 
action to generate the vanishing ideal $I(Z)$, $Z=J^{-1}(0)$, it is necessary and 
sufficient, that the following \emph{nonpositivity condition} applies:
for all $\xi\in \mathfrak g$ and $z\in Z$ one has either
\begin{enumerate}
\item $J(\xi)=0$ in a neighborhood $U\subset M$ of $z$, or
\item in every neighborhood $U\subset M$ of $z$ the function $J(\xi)$ takes strictly 
      positive as well as strictly negative values.
\end{enumerate}  
This nonpositivity condition and Theorem \ref{acyckrit} make it easy to provide
first nontrivial examples.
\subsubsection{Zero Angular Momentum for $m$ particles in $\mathbb R^2$.} We consider 
the system of $m$ particles in $\mathbb R^2$ with zero total angular momentum 
(see e.g.~\cite[Section 5]{LerMontSj} and \cite[Section 6]{HuebschDesc}). More 
precisely, the phase space is $M:=(T^*\mathbb R^2)^m$ and we let 
$SO(2,\mathbb R)\cong S^1$ act on it by lifting the diagonal action, i.e.,
\begin{eqnarray*}
  SO(2)\times M&\to& M\\
  (g,(\q_1,\p^1,\dots,\q_m,\p^m))&\mapsto& (g\q_1,g\p^1,\dots,g\q_m,g\p^m),
\end{eqnarray*} 
where $\q_i=(q^1_i,q^2_i)^t$ and $\p^i=(p_1^i,p_2^i)^t$ for $i=1,\dots,m$.
The moment map $J:M\to\mathfrak{so}(2)=\mathbb R$ is given by 
$J(\q,\p)=\sum_{i=1}^m q^1_ip_2^i-q^2_ip_1^i$. In \cite{LerMontSj} the reduced space 
is described as a branched double cover of the closure of a certain coadjoint orbit 
of $\mathfrak{sp}(m,\mathbb R)$. The moment map $J$ obviously satisfies the 
nonpositivity condition above. Since $Z=J^{-1}(0)$ is of codimension 1, this implies 
that the Koszul complex (cf.~Section \ref{KosSect}) is a resolution of 
$\mathcal C^{\infty}(Z)$.

\subsubsection{An $S^1$-action with a worse-than-orbifold quotient.}
The following example is taken from \cite[p.125]{CushSjam}. Consider the $S^1$-action 
on $\mathbb C^4$, endowed with symplectic form 
$\omega=\frac{i}{2}\sum_kdz_k\wedge d\bar z_k$, given by 
$\mathrm e^{i\vartheta}\cdot(z_1,z_2,z_3,z_4):=
(\mathrm e^{\mathrm i\vartheta}z_1,\mathrm e^{\mathrm i\vartheta}z_2,
\mathrm e^{-\mathrm i\vartheta}z_3,\mathrm e^{-\mathrm i\vartheta}z_4)$. 
The moment map for the action is
\[J(z_1,z_2,z_3,z_4)=\frac{1}{2}(|z_3|^2+|z_4|^2-|z_1|^2-|z_2|^2).\]
The constraint surface $Z$ is the real cone $C(S^3\times S^3)$, and by a topological 
argument (see \cite{CushSjam}), the reduced space $C(S^3\times_{S^1}S^3)$ can not be 
an orbifold.
Since $J$ clearly satisfies the nonpositivity condition above,
it generates the vanishing ideal $I(Z)$. Again, we conclude that the Koszul complex 
is a resolution of $\mathcal C^{\infty}(Z)$.
\subsubsection{A $T^2$-action on $\mathbb C^4$.} We consider example 7.7 from 
\cite{AGJ}. The action is given by 
$T^2\times \mathbb C^4\to \mathbb C^4$, $((\vartheta_1,\vartheta_2),(z_1,z_2,z_3,z_4))
 \mapsto (\mathrm e^{\mathrm i (\alpha  \vartheta_1+\beta\vartheta_2)}z_1, 
 \mathrm e^{-\mathrm i\vartheta_2}z_2,\mathrm e^{\mathrm i\vartheta_1}z_3,
 \mathrm e^{-\mathrm i\vartheta_2}z_4)$ for $\alpha,\beta \in \mathbb Z$. 
A moment map for the action is 
$J:\mathbb C^4\to \mathbb R^2$, $J(z_1,z_2,z_3,z_4):=
\frac {1}{2}(-\alpha|z_1|^2-|z_3|^2,-\beta|z_1|^2+|z_2|^2-|z_4|^2)$. 
$J$ satisfies the nonpositivity condition for $\alpha<0$. An elementary calculation 
gives that also condition b) of Theorem \ref{acyckrit} is true. Consequently, the 
corresponding Koszul complex is a resolution of the space of smooth functions on 
$Z:=J^{-1}(0)$.

\subsection{Hamiltonian actions of nonabelian Lie groups}

As the nonpositivity condition, in the case of nonabelian group actions, is only 
\emph{necessary} for the ideal $I(Z)\subset \mathcal C^{\infty}(M)$ to be generated 
by $J_1,\dots ,J_\ell$, the reasoning here is usually more intricate. In \cite{AGJ} 
it was proven that the latter is the case iff in every normal coordinate system the 
ideal $I$ generated by the moment map in the real polynomial ring 
$\mathbb R[x^1,\dots,x^{2n}]$ is \emph{real} in the sense of real algebraic geometry 
(cf.~\cite[Theorem 6.3]{AGJ}). Recall that an ideal $I$ in 
$\mathbb R[x^1,\dots, x^m]$ is real, if it coincides with its \emph{real radical} 
\begin{displaymath}
\begin{split}
  \sqrt[\mathbb R]{I} & := \\
  & \big\{f\in\mathbb R[x^1,\dots, x^m] \mid f^{2i}+\sum_{j=0}^k g_j^2
  \in I\text{ for some $i$ and 
  $g_1,\dots , g_k\in \mathbb R[x^1,\dots, x^m]$}\big\}.
\end{split}
\end{displaymath} 
In \cite{AGJ} we find the following criterion for such an ideal to be real. 
\begin{theorem} 
\label{GotayThm} 
Let $I$ be an ideal in $\mathbb R[x^1,\dots, x^m]$. Then $I$ is real, if and only 
if the following two conditions hold:
\begin{enumerate}
\item 
  $I_\mathbb C:=I\otimes_\mathbb R \mathbb C$ is radical in 
  $\mathbb C[x^1,\dots, x^m]$, and
\item 
 for every irreducible component $W\subset \mathbb C^m$ of the (complex) locus of 
 $I_\mathbb C$ \[\dim_{\mathbb R} (W\cap \mathbb R^m)=dim_{\mathbb C}(W).\]
\end{enumerate}
\end{theorem} 
In other words, in order to know whether the ideal $I$ is real, it is enough to gain 
detailed insight into the complex algebraic geometry behind the scene (e.g.~knowing 
the primary decomposition of $I_\mathbb C$). Regardless the fact that the varieties in 
question are cones, there is no straightforward way to provide this information. A 
basic example, which one is tempted to consider is zero angular momentum of one 
particle in dimension $n$. Since the components of the moment map can be written as 
the $2\times2$-minors of a $2\times n$-matrix, the ideal $I_\mathbb C$ is prime, and 
the complex locus is of dimension $n+1$ by a theorem of Hochster \cite{Hochster}. It 
follows easily, that the ideal $I$ is real. Unfortunately, this example is not a 
complete intersection for $n\ge3$. The only class of nonabelian examples, which the 
authors are aware of, where generating and complete intersection hypothesis are true 
at the same time, is the following. 
\subsubsection{Commuting Varieties.} 
Let $S$ the space of symmetric $n\times n$-matrices with real entries. We let 
$SO(n)$ act on $S$ by conjugation and we lift this action to an action of $SO(n)$ on 
the cotangent bundle $T^*S=S\times S$. This action is hamiltonian with the moment map
\begin{eqnarray*}
  J:S\times S&\to&\wedge^2\mathbb R^n=\mathfrak{so}(n)^*\\
  (Q,P)&\mapsto&[Q,P],
\end{eqnarray*}
where we have identified $\mathfrak{so}(n)^*$ with the space $\wedge^2\mathbb R^n$ of 
antisymmetric $n\times n$-matrices. The complex locus $Z_{\mathbb C}$ defined by the 
these $\frac{1}{2}n(n-1)$ quadratic equations is an instance of what is called a 
\emph{commuting variety}. In \cite{BrenPiVas} it was shown that $Z_{\mathbb C}$ is 
irreducible of codimension $\frac{1}{2}n(n-1)$, and the ideal generated by the 
coefficients of $J$ in the complex polynomial ring is prime. Let 
$S_{\mathrm{reg}}\subset S$ be the open subset of symmetric matrices with pairwise 
distinct eigenvalues. Since the action of $SO(n)$ on $T^*S_{\mathrm{reg}}$ is locally 
free, it follows that $Z\cap T^*S_{\mathrm{reg}}$ is of codimension $\frac{1}{2}n(n-1)$ 
likewise. As a consequence of Theorem \ref{GotayThm}, the components of $J$ generate
the vanishing ideal $I(Z)$ in $\mathcal C^{\infty}(T^*S)$. It is easy to see, that 
$T_zJ$ is surjective for $z\in Z\cap T^*S_{\mathrm{reg}}$. By Theorem \ref{acyckrit} 
below, the Koszul complex is a resolution of the space of smooth function on $Z$. 
Using invariant theory, the reduced space was identified in \cite{LerMontSj} as the 
quotient $(\mathbb R^n \times \mathbb R^n)/S_n$, the symmetric group $S_n$ acting 
diagonally. Note that the results of \cite{BrenPiVas} have been generalized to  
moment maps of the isotropy representations of symmetric spaces of maximal rank 
\cite{Pany}.

\section{Koszul resolution}\label{KosSect}
Given a smooth map $J:M\to \mathbb R^\ell=:V^*$ we consider the Koszul holomogical 
complex of the sequence  of ring elements $J_1,\dots, J_\ell\in \mathcal C^{\infty}(M)$,
but we will view it later artificially as a cochain complex. In other words, we define 
the space of (co)chains to be $K^{i}:=K_{-i}(M,J):=S^i_{\mathcal C^{\infty}(M)}(V[1])$,
i.e.~the free (super)symmetric $\mathcal C^{\infty}(M)$-algebra generated by the 
graded vector space $V[1]$, where we consider $V$ to be concentrated in degree zero. 
$K_{\bullet}$ may also be viewed as the space of sections of the trivial vector bundle 
over $M$ with fibre $\wedge^\bullet V$. Denoting by $e^1,\dots, e^\ell$ the canonical 
bases of the dual space $V$ of $V^*=\mathbb R^\ell$, we define the Koszul differential 
$\partial:=\sum_a J_a i(e^a)$, where the $i(e^a)$ are the derivations extending the 
dual pairing. We will say, in accordance with \cite{Bourb10}, that 
$J_1,\dots, J_\ell\in \mathcal C^{\infty}(M)$ is a \emph{complete intersection}, if 
the homology of the Koszul complex vanishes in degree $\ne 0$.

Now we would like to have a simple geometric criterion for $J$ to be a complete 
intersection. We achieve this goal only after knowing that $J$ generates the vanishing 
ideal (which is sometimes difficult to decide).
\begin{theorem} \label{acyckrit} Let M be an analytic manifold and 
$J:M\to \mathbb R^\ell$  an  analytic map, such that the following conditions are true
\begin{enumerate}
\item $(J_1,\dots,J_\ell)$ generate the vanishing ideal of $Z:=J^{-1}(0)$ in 
      $\mathcal C^\infty(M)$,
\item the regular stratum $Z_r:=\{z\in Z\:|\: T_zJ \mbox{ is surjective}\}$ is dense 
      in $Z:=J^{-1}(0)$.
\end{enumerate}
      Then the Koszul complex $K:=K(M,J)$ is acyclic and $H_0=\mathcal C^\infty(Z)$.
\end{theorem}
\begin{proof} We will show that the Koszul complex $K(\mathcal C^\omega_x(M),J) $ is 
acyclic for the ring $\mathcal C^\omega_x(M)$ of germs in $x$ of real analytic 
functions. Then it will follow that the Koszul complex $K(\mathcal C^{\infty}(M),J)$ 
is acyclic, since the ring of germs of smooth functions $\mathcal C^\infty_x(M)$ is 
flat over $\mathcal C^\omega_x(M)$ (see \cite[p.118]{Toug}), and the sheaf of smooth 
functions on $M$ is fine. Since $\mathcal C^\omega_x(M)$ is noetherian, Krull's 
intersection theorem says that $\cap_{r\ge0} I^r_x=0$, where $I_x$ is the ideal of 
germs of analytic functions vanishing on $Z$. According to \cite[A X.160]{Bourb10}, 
it is therefore sufficient to show that  $H_1(\mathcal C^\omega_x(M),J)=0$. Note that 
since $J$ generates the vanishing ideal of $Z$ in $\mathcal C^\infty(M)$, it also 
generates the vanishing ideal of $Z$ in $\mathcal C^\omega_x(M)$. This can easily be 
seen using M. Artin's approximation theorem (see e.g.\cite{Ruiz}). Suppose 
$f=\sum_a f^a\:e_a\in K_1$ is a cycle, i.e. $\partial f=\sum_a J_af^a=0$. Since the 
restriction to $Z$ of the Jacobi matrix $D(\sum_a J_af^a)$ vanishes,  we conclude 
(using condition b)) that $f^a_{|Z}=0$ for all $a=1,\dots,\ell$. Since $J$ generates 
the vanishing ideal, we find an $\ell\times\ell$-matrix $F= (F^{ab})$ with smooth 
(resp. analytic) entries such that $f^a=\sum_b F^{ab}J_b$. It remains to be shown, that 
this matrix can be choosen to be \emph{antisymmetric}. We have to distinguish two 
cases.  If $x\notin Z$, the claim is obvious, since then one can take for example 
$F^{ab}:=(\sum_a J_a^2)^{-1}(J_bf^a-J_a f^b)$.
So let us consider the other case  $x \in Z$. We then introduce some formalism
to avoid tedious symmetrization arguments. 
Let $E$ denote the free $k:=\mathcal C^\omega_x(M)$-module on $\ell$ generators, 
and consider the Koszul-type complex $SE\otimes \wedge E$. Generators of the symmetric part 
will be denoted by $\mu_1,\dots,\mu_\ell$, generators of the Grassmann part by 
$e_1,\dots,e_\ell$, respectively. We have two derivations 
$\delta:=\sum_a e_a\wedge\frac{\partial}{\partial \mu_a}:S^nE\otimes \wedge^m E\to 
S^{n-1}E\otimes \wedge^{m+1}E$, and 
$\delta^*:=\sum_a \mu_a i(e^a):S^nE\otimes \wedge^m E\to S^{n+1}E\otimes \wedge^{m-1}E$. 
They satisfy the well known identities: $\delta^2=0$, $(\delta^*)^2=0$ and 
$\delta\delta^*+\delta^*\delta=(m+n)\id$. Furthermore, we introduce the two commuting 
derivations $i_J:=\sum_a J_a i(e^a)$ and $d_J=\sum_a J_a\frac{\partial}{\partial \mu_a}$. 
They obey the identities $i_J^2=0$, $[i_J,\delta]=d_J$, $[d_J,\delta^*]=i_J$ and 
$[i_J,\delta^*]=0=[d_J,\delta]$. We interprete the cycle $f$ above as being in $E\otimes k$ 
and the matrix $F$ as a member of $E\otimes E$. We already know that $d_Jf=0$ implies $f=i_JF$. 
This argument may be generalized as follows: if $a\in S^nE\otimes k$ obeys $d_J^na=0$, then 
there is an $A\in S^nE\otimes E$ such that $a=i_J A$. The proof is easily provided by taking 
all $n$-fold partial derivatives of $d_J^na=0$, evaluating the result on $Z$ and using 
conditions a) and b).
We now claim that there is a sequence of $F_{(n)}\in S^{n+1}E\otimes E$, $n\ge 0$, such that 
$F=F_{(0)}$, $\delta^*F_{(n)}=(n+2)i_JF_{(n+1)}$ and 
\begin{eqnarray}\label{nteGl}
 f=d_J^ni_JF_{(n)}+i_J\delta^*\underbrace{\Big(\sum_{i=0}^{n-1}\frac{1}{i+2}\:d_J^i\delta 
 F_{(i)}\Big)}_{=:B_{n-1}}\quad\mbox{ for all }n\ge1.
\end{eqnarray} 
We prove this by induction. Setting $B_{-1}:=0$, we may start the induction with $n=0$, where 
nothing has to be done. Suppose now, that the claim is true for $F_{(0)},\dots ,F_{(n)}$. 
We obtain $f=\frac{1}{n+2}d_J^ni_J\big(\delta \delta^*F_{(n)}+\delta^*\delta F_{(n)}\big)+
i_J\delta^*B_{n-1}=\frac{1}{n+2}d_J^{n+1}\delta^*F_{(n)}+i_J\delta^*B_n$,
where we made use of the relations $[d^n_Ji_J,\delta^*]=0$ and $[d^n_Ji_J,\delta]=d^{n+1}_J$. 
Since $0=d_Jf=d_J^{n+2}\delta^*F_{(n)}$, we find an $F_{(n+1)}$ such that 
$\frac{1}{n+2}\delta^*F_{(n)}=i_JF_{(n+1)}$, and the claim is proven. 
Finally, we want to take the limit of equation (\ref{nteGl}) 
as $n$ goes to $\infty$. 
For this limit to make sense, we have to change the ring to the ring of formal power series. 
Let us denote this change of rings by 
$\hat{}:\mathcal C_x^\omega(M)\to\mathbb K[[x^1,\dots,x^n]]$. 
Since by Krull's intersection theorem $\cap_{r\ge0} \hat{I}^r=0$ ($\hat I$ the ideal generated by 
$\hat J_1,\dots,\hat J_{\ell}$), we obtain a formal solution of the problem: 
$\hat f=i_{\hat J}\delta^*B_{\infty}$, where $B_{\infty}:=\sum_{i=0}^{\infty}\frac{1}{i+2}\:
d_{\hat J}^i\delta \hat F_{(i)}$ is well defined since $\hat I$ contains the maximal ideal. 
Applying M. Artin's approximation theorem yields an analytic solution, and we are done. 
\end{proof}

The above reasoning can be considered to be folklore, as the subtlety of finding an 
\emph{antisymmetric} source term is often swept under the rug in semirigorous arguments. 
The next theorem though is a consequence of rather 
deep analytic results. The problem of splitting the Koszul resolution in the context of
Fr\'echet spaces was also addressed in \cite{DomJac} from a different perspective.
\begin{theorem} 
  Let $M$ be a smooth manifold, $J:M\to \mathbb R^\ell$ be smooth map such that around every 
  $m\in M$ there is a local chart in which $J$ is real analytic. Moreover, assume that the 
  Koszul complex $K=K(M,J)$ is a resolution of $\mathcal C^\infty(Z)$, $Z=J^{-1}(0)$. Then 
  there are a prolongation map 
  $\prol: \mathcal C^{\infty}(Z)\to \mathcal C^{\infty}(M)$ 
  and contracting homotopies $h_i: K_i\to K_{i+1}$, $i\ge 0$, 
  which are continuous in the respective Fr\'echet topologies, such that
\begin{eqnarray}
\label{KosContr}
  \big(\mathcal C^{\infty}(Z),0\big)\:
  \begin{array}{c}{\res}\\\leftrightarrows\\{\prol}
  \end{array}\:\big(K,\partial\big),h
\end{eqnarray}
  is a contraction, i.e.~$\res$ and $\prol$ are chain maps and $\res\:\prol=\id$ and 
  $\id-\prol\:\res=\partial h+h\partial$. If necessary, these can be adjusted in 
  such a way, that the side conditions (see Appendix \ref{HPT})
  $h_0\:\prol=0$ and 
  $h_{i+1}\:h_i=0$
  are fulfilled. If, moreover, a compact Lie group $G$ acts smoothly
  on $M$, $G$ is represented on $\mathbb R^\ell$ and $J:M\to \mathbb R^\ell$ is equivariant, 
  then $\prol$ and $h$ can additionally be chosen to be equivariant.
\end{theorem}
\begin{proof} A closed subset $X\subset\mathbb R^n$ is defined to have the \emph{extension property}, if there is a continuous linear map $\lambda:\mathcal C^\infty(X)\to \mathcal C^\infty(\mathbb R^n)$, such that $\res\: \lambda=\id$. 
The extension theorem of E. Bierstone and G. W. Schwarz, \cite[Theorem 0.2.1]{Schwarzbier} says that Nash subanalytic sets (and hence closed analytic sets) have the extension property. 
Using a partition of unity, we get a continuous linear map $\lambda:\mathcal C^\infty(Z)\to \mathcal C^\infty(M)$, such that $\res\: \lambda=\id$. 
In the same reference, one finds a ``division theorem'' (Theorem 0.1.3.), which says that for a matrix $\varphi\in \mathcal C^\omega(\mathbb R^n)^{r,s}$ of analytic functions the image of $\varphi:\mathcal C^\infty(\mathbb R^n)^s\to C^\infty(\mathbb R^n)^r$ is closed, and there is a continuous split $\sigma:\mathrm{im}\:\varphi\to\mathcal C^\infty(\mathbb R^n)^s$ such that $\varphi\:\sigma=\id$. 
Using a partition of unity, we conclude that there are linear continuous splits $\sigma_i:\mathrm{im}\:\partial_{i+1}\to K_{i+1}$ for the Koszul differentials $\partial_{i+1}:K_{i+1}\to K_i$ for $i\ge 0$, i.e. $\partial_{i+1}\:\sigma_i=\id$. We observe that $\mathrm{im}\:\lambda\oplus\mathrm{im}\:\partial_1=K_0$, since for every $x\in K_0$ the difference $x-\lambda\:\res x$ is a boundary due to exactness and the sum is apparantly direct. 
Similarly, we show that $\mathrm{im}\:\sigma_i\oplus\mathrm{im}\:\partial_{i+2}=K_{i+1}$ for $i\ge 0$. The next step is to show that $\mathrm{im}\:\sigma_i$ is a \emph{closed} subspace of $K_0$. Therefor we assume that $(x_n)_{n\in\mathbb N}$ is a sequence in $\mathrm{im}\:\partial_{i+1}$ such that $\sigma_i(x_n)$ converges to $y\in K_{i+1}$. Then $x_n=\partial_{i+1}\sigma_i(x_n)$ converges to $\partial_{i+1} y$, since $\partial_{i+1}$ is continuous. Since $\partial_{i+1} y$ is in the domain of $\sigma_i$, we obtain that $\sigma_i(x_n)$ converges to $\sigma_i\partial_{i+1} y=y\in \mathrm{im}\:\sigma_i$. Similarly, we have that $\mathrm{im}\:\lambda$ is a closed subspace of $K_0$. Altogether, it is feasible to extend $\sigma_i$ to a linear continuous map $K_i\to K_{i+1}$ (cf. \cite[ p.133]{Rudin}). If necessary, $\lambda$ and $\sigma_i$ can be made equivariant by averaging over $G$, since $\res$ and $\partial$ are equivariant. We observe that we have 
$\lambda \res_{|\mathrm{im}\lambda}=\id$ and 
$\lambda \res_{|\mathrm{im}\partial_1}=0$ and analogous equations in higher degrees. 
We now replace $\lambda$ by $\prol:=\lambda-\partial_1\sigma_0 \lambda$ and $\sigma_i$ 
by $h_i:=\sigma_i-\partial_{i+2}\sigma_{i+1}\sigma_i$ for $i\ge0$. These maps share 
all of the above mentioned properties with $\lambda$ and $\sigma_i$. Additionally, we 
have ${\partial_1 h_0}_{|\mathrm{im}(\prol)}=0$ and 
${\partial_{i+2}h_{i+1}}_{|\mathrm{im}(h_i)}=0$ for $i\ge0$. This concludes the 
construction of (\ref{KosContr}). The side conditions can be achieved by algebraic 
manipulations (see Appendix \ref{HPT}).  Note that these modifications do not ruin 
the equivariance.
\end{proof}

A crucial property of the Koszul resolution is that it is a differential graded commutative algebra. In the present context, where the constraint functions are the 
components of a moment map, it has the following extra feature. The Lie 
algebra $\mathfrak g$ acts on it by even derivations, extending the actions 
on $\mathfrak g$ and on $\mathcal C^{\infty}(M)$.

\section{Classical homological reduction} \label{classred}

The BRST-algebra is defined to be 
$\mathscr A:=S_{\mathcal C^{\infty}(M)}\big(\mathfrak g[1]\oplus \mathfrak g^*[-1]\big)$, 
i.e. the free graded commutative $\mathcal C^{\infty}(M)$-algebra 
generated by $\mathfrak g$ (of degree $-1$) and $\mathfrak g^*$ (of 
degree $1$). We adopt the usual convention to call the elements of $\mathfrak g^*$ and $\mathfrak g$  \emph{ghosts} and \emph{antighosts}, respectively. We will frequently refer to a basis $e_1,\dots ,e_\ell$ and $e^1,\dots ,e^\ell$ of $\mathfrak g$ and  $\mathfrak g^*$, respectively (we will use latin indices: $a,b,\dots$). There is an even graded Poisson bracket on $\mathscr A$ 
extending that on $M$, which is uniquely defined by the requirements 
$\{\alpha,x\}=\langle\alpha,x\rangle$ and $\{f,x\}=0=\{f,\alpha\}$ for 
all $x\in\mathfrak g$, $\alpha \in \mathfrak g^*$ and 
$f\in\mathcal C^{\infty}(M)$. With the Lie bracket and the moment map 
we build an element 
$\theta:= -\frac{1}{4}\sum_{a,b,c}\:f_{ab}^c\:e^ae^be_c+\sum_a J_a e^a\in\mathscr A^1$, where the $f_{ab}^c$ 
are the structure constants of $\mathfrak g$. An easy calculation yields 
$\{\theta,\theta\}=0$, hence $\mathscr D:=\{\theta,?\}$ is a differential. 
Summing up, we obtain a differential graded Poisson algebra 
$(\mathscr A,\{,\},\mathscr D=\{\theta,?\})$, 
we call $\theta$ the \emph{BRST-charge} and $\mathscr D$ 
the \emph{classical BRST-differential}. 

Closer examination shows that $\mathscr D=\delta+2\partial$ is a linear 
combination of two supercommuting differentials. Here, $\delta$ is 
the codifferential 
of the Lie algebra cohomology corresponding to the $\mathfrak g$-module 
$S_{\mathcal C^{\infty}(M)}(\mathfrak g[1])$, this representation will be denoted by 
$\mathrm L$, and $\partial=\sum_a J_ai^a$ is 
the extension of the Koszul differential. We view $\mathscr D$ as a perturbation 
(see Appendix \ref{HPT}) of the acyclic differential $\partial$.
 
We extend the restriction map $\res$ to a map 
$\res:\mathscr A\to S_{\mathcal C^\infty(Z)}(\mathfrak g^*[-1])$ by setting it zero for all 
terms containing antighosts and restricting the coefficients. In the same fashion, we extend 
$\prol$ to a map $S_{\mathcal C^\infty(Z)}(\mathfrak g^*[-1])\to\mathscr A$ extending the 
coefficients. 

Since the moment map $J$ is $G$-equivariant, $G$  acts on $Z=J^{-1}(0)$. Hence 
$\mathcal C^\infty(Z)$ is a $\mathfrak g$-module, this representation will be denoted by 
$\mathrm{L}^z$. Note that $\mathrm{L}^z_X=\res\;\mathrm L_X\;\prol$ for all $X\in\mathfrak g$. 
We identify  $S_{\mathcal C^\infty(Z)}(\mathfrak g^*[-1])$  with the space of cochains of Lie 
algebra cohomology $C^\bullet\big(\mathfrak g,\mathcal C^\infty(Z)\big)$.
Let us denote 
$d:C^\bullet\big(\mathfrak g,\mathcal C^\infty(Z)\big)\to C^{\bullet+1}
\big(\mathfrak g,\mathcal C^\infty(Z)\big)$ the codifferential of Lie algebra cohomology 
coresponding to $\mathrm{L}^z$. Since $\res$ is a morphism of $\mathfrak g$-modules we obtain 
$d\;\res=\res\;\delta$.

\begin{theorem}There are $\mathbb K$-linear maps 
$\Phi:C^\bullet\big(\mathfrak g, \mathcal C^{\infty}(Z)\big)\to \mathscr A^\bullet$
and $H: \mathscr{A}^\bullet\to \mathscr{A}^{\bullet-1}$ which are continuous in the 
respective Fr\'echet topologies such that
\begin{eqnarray}\label{cbrstSDR}
\Big(C^\bullet\big(\mathfrak g, \mathcal C^{\infty}(Z)\big),d\Big)\:
\begin{array}{c} {\res}\\\leftrightarrows\\ {\Phi}
\end{array}
\:(\mathscr A^\bullet,\mathscr D),H
\end{eqnarray}
is a contraction.
\end{theorem}
\begin{proof} Apply lemma \ref{BPL1} to the perturbation $\qbrst$ of $2\partial$. Explicitly, 
we get $H:=\frac{1} {2} h\sum_{j=0}^{\ell}(-\frac{1}{2})^j(h\delta+\delta h)^j $ and 
$\Phi=\prol-H(\delta\;\prol-\prol\;d)$, which are obviously Fr\'echet continuous. 
Note that from $h\prol=0$ and $h^2=0$ it follows that $H\Phi=0$ and $H^2=0$. If $\prol$ is 
chosen to be equivariant, then the expression for $\Phi$ simplifies to $\Phi=\prol$. In the 
same way one gets $H=\frac{1}{2}h$, if $h$ is equivariant.
\end{proof}

\begin{corollary} There is a  graded Poisson structure on 
$\mathrm H^\bullet\big(\mathfrak g,\mathcal C^\infty(Z)\big)$. If $[a],[b]$ are the cohomology 
classes of $a,b\in C^\bullet\big(\mathfrak g, \mathcal C^{\infty}(Z)\big)$, then the bracket is 
given given by $\{[a],[b]\}:=[\res\{\Phi(a),\Phi(b)\}]$. The restriction of this bracket to 
$\mathrm H^0\big(\mathfrak g,\mathcal C^\infty(Z)\big)=\mathcal C^\infty (Z)^{\mathfrak g}$ 
coincides with the Dirac reduced Poisson structure. 
\end{corollary}
\section{The quantum BRST-algebra}\label{quantumBRST}
In this section we will introduce the quantum BRST algebra, which is 
$\mathbb K[[\nu]]$-dg algebra $(\mathscr{A}^\bullet[[\nu]],*,
\mathscr D_{\nu})$ deforming the classical dg Poisson algebra 
$(\mathscr A^\bullet,\{,\},\mathscr D)$. The exposition parallels that 
of \cite{BHW}. In order to define a graded product $*$ on $\mathscr A[[\nu]]$, 
we use on one hand a Clifford multiplication 
$x\cdot y:=\mu\big(\operatorname e^{-2\nu\: \sum_a i^a\otimes i_a}(x\otimes y)\big)$ 
for $x,y\in S_{\mathbb K}(\mathfrak g[1]\oplus \mathfrak g[-1])$.
Here $\mu$ denotes the supercommutative multiplication, $i^a$ and $i_a$ are the  left derivations extending the dual 
pairing with $e^a$ and $e_a$, respectively and $\otimes$ denotes the graded tensor product. 
On the other hand, we will need a star product $\star$ on $M$, 
which is compatible with the $\mathfrak g$-action in the following sense
\begin{eqnarray}\label{qmomentmap} 
\mathbb J(X)\star \mathbb J(Y)-\mathbb J(Y)\star 
\mathbb J(X)=\nu\mathbb J([X,Y])\mbox{ for all } X,Y\in \mathfrak g,
\end{eqnarray}
where $\mathbb J=J+\sum_{i\ge1}\nu^i J_{(i)}\in \mathscr A^1[[\nu]]$
is a deformation of the moment map $J$. In other words, $\star$ is 
\emph{quantum  covariant} for the \emph{quantum moment map} $\mathbb J$. 
For $f,g\in \mathcal C^{\infty}(M)$ and 
$x,y \in S\big(\mathfrak g[1]\oplus \mathfrak g^*[-1]\big)$ we define 
$(fx)*(gy):=(f\star g)\: (x\cdot y)$. 
Note that $*$ is graded. The next step is to quantize the BRST-charge. 
Luckily, we are done with (see e.g. \cite{KostStern})
\[\theta_\nu:= -\frac{1}{4}\sum_{a,b,c}\:f_{ab}^c\:e^ae^be_c+
\mathbb \sum_a\mathbb J_a\:e^a+\frac{1}{2}\nu\sum_a f_{ab}^be^a\:\in\mathscr A^1[[\nu]],\] 
since a straightforward calculation yields $\theta_\nu*\theta_\nu=0$. 
We define the \emph{quantum BRST differential} to be 
$\qbrst:=\frac{1}{\nu}\ad_*(\theta_\nu)$.
 
Before we take a closer look, at $\qbrst$ let us introduce some terminology. We define the 
(superdifferential) operators $\qce,\mathscr R,q,u:\mathscr A^\bullet\to\mathscr A^{\bullet+1},$
\begin{eqnarray*}
\qce(f)&:=&-\frac{1}{2}\sum_{a,b,c}f_{ab}^c\:e^ae^b\:i_c(f)+\sum_{a,b,c}f_{ab}^c\:e^ae_c\:i_b(f)+
\sum_a\:e^a\:\frac{ 1}{\nu}[\mathbb J_a,f]_*, \\      
\mathscr R(f)&:=&\sum_ai^a\:f * \mathbb J_a,\quad\quad\quad\quad\quad\quad\quad
\mbox{``right multiplication''}\\
q(f)&:=&-\frac{1}{2}\sum_{a,b,c}f_{ab}^c\:e_c\:i^ai^b(f),\quad\quad\quad\quad\quad\quad
\mbox {``quadratic ...''} \\
u(f)&:=&\sum_{a,b}f_{ab}^b\:i^a(f),\quad\quad\quad\quad\quad\quad \quad\quad
\mbox{``unimodular term''}
\end{eqnarray*} 
for $f\in \mathscr A$. Note that $\qce$ is the coboundary operator of Lie algebra cohomology 
corresponding to the representation 
\begin{eqnarray}
\mathbb L_X:S_{\mathcal C^\infty(M)}(\mathfrak g[1])[[\nu]]&\to& S_{\mathcal C^\infty(M)}
(\mathfrak g[1])[[\nu]],\nonumber \\
af&\mapsto& (\ad_X(a))f +a\:\nu^{-1}(\mathbb J(X)\star f- f\star \mathbb J(X)),
\end{eqnarray} 
where $X\in \mathfrak g$, $a\in S_{\mathbb K}(\mathfrak g[1])$ and 
$f\in \mathcal C^\infty(M)[[\nu]]$. Finally, we set
\[\qkos:=\mathscr R+\nu\Big(\frac{1}{2}u-q\Big).\]
This operator will be called the \emph{deformed} or \emph{quantum Koszul differential}. 
Note that $\qkos$ is a homomorphism of $\mathcal C^\infty(M)[[\nu]]$-left-modules. As a side 
remark, $\qkos$ may also be interpreted as a differential of Lie algebra homology of a certain 
representaion of $\mathfrak g$. This point of view was adopted in \cite{Sevost}. 
\begin{theorem} \label{doppelcompl}The quantum BRST differential $\qbrst=\qce+2\qkos$ is a 
linear combination of two supercommuting differentials $\qce$ and $\qkos$. 
\end{theorem}
\begin{proof} Straightforward calculation.
\end{proof}

\section{Quantum reduction} \label{quantumred}

The main idea, which we follow  in order to compute the quantum BRST cohomology (i.e.~the 
cohomology of $(\mathscr A[[\nu]],\qbrst)$), is to provide a deformed version of the 
contraction (\ref{cbrstSDR}). This will be done by applying Lemma \ref{BPL2} to the 
contraction (\ref{KosContr}) for the perturbation $\qkos$ of $\partial$ and then applying 
Lemma \ref{BPL1} for the perturbation $\qbrst$ of $2\qkos$. We will also need to examine a 
deformed version of the representation $\mathrm{L}^z$ of $\mathfrak g$ on $\mathcal C^\infty(Z)$. 

\begin{proposition} If we choose $h_0$ such that $h_0\prol=0$, then there are deformations of 
the restriction map 
$\res_\nu=\res+\sum_{i\ge 1}\nu^i\;\res_{i}:\mathcal C^{\infty}(M)\to 
\mathcal C^{\infty}(Z)[[\nu]]$ and of the contracting homotopies 
$\qhop_i=h_i+\sum_{j\ge 1}\nu^j\;h_i^j:K_i[[\nu]]\to K_{i+1}[[\nu]]$,
which are a formal power series of Fr\'echet continuous maps and such that 
\begin{eqnarray}
\label{qKosContr}
\big(\mathcal C^{\infty}(Z)[[\nu]],0\big)\:\begin{array}{c} {\res_\nu}\\\leftrightarrows\\ {\prol}
\end{array}\:\big(K[[\nu]],\qkos\big),\qhop
\end{eqnarray}
is a contraction with $\qhop_0\;\prol=0$. Explicitly, we have 
\[\res_\nu:=\res\;(\id+(\qkos_1 -\partial_1)h_0)^{-1}.\]
If we choose $h$ to be $\mathfrak g$-equivariant, the same is true for $\qhop$.
\end{proposition}
\begin{proof} Apply lemma \ref{BPL2} to the perturbation $\qkos$ of $\partial$.
\end{proof}

We now define the quantized representation $\mathbb L^z$ of $\mathfrak g$ on 
$\mathcal C^\infty (Z)[[\nu]]$ by setting 
\[\mathbb L^z_X:=\res_\nu\;\mathbb L_X \;\prol \quad\mbox{ for $X\in \mathfrak g$.}\]
That this is in fact a representation, follows easily from the observation 
$\mathbb L_X\qkos-\qkos\mathbb L_X=0$ for all $X\in \mathfrak g$ (this is a consequence of 
Theorem \ref{doppelcompl}), and from $\qhop$ being a contracting homotopy. In the same fashion 
as in Section \ref{classred}, we define 
$\qdz:C^\bullet(\mathfrak g,\mathcal C^\infty (Z)[[\nu]])\to C^{\bullet+1}
(\mathfrak g,\mathcal C^\infty (Z)[[\nu]])$ to be the differential of Lie algebra cohomology 
of the representation $\mathbb L^z$, i.e. $\qdz\;\res_\nu=\res_\nu\;\qce$. 
In the same manner, we extend $\res_\nu$ and $\qhop$  as in Section \ref{classred} to maps 
$\res_\nu:\mathscr A\to C(\mathfrak g,\mathcal C^{\infty}(Z)[[\nu]]\big)$ and 
$\qhop:\mathscr A^\bullet[[\nu]]\to\mathscr A^{\bullet-1}[[\nu]]$. 

\begin{theorem} 
\label{Hauptsatz}
  There are $\mathbb K[[\nu]]$-linear maps 
  $\Phi_\nu:C^\bullet\big(\mathfrak g, \mathcal C^{\infty}(Z)\big)\to \mathscr A^\bullet[[\nu]]$ 
  and $H_\nu: \mathscr{A}^\bullet\to \mathscr{A}^{\bullet-1}[[\nu]]$,
  which are series of  Fr\'echet continuous maps such that
\begin{eqnarray*}
  \Big(C^\bullet\big(\mathfrak g, \mathcal C^{\infty}(Z)[[\nu]]\big),\qdz\Big)\:
\begin{array}{c} {\res_\nu}\\\leftrightarrows\\ {\Phi_\nu}
\end{array}
  \:(\mathscr A^\bullet[[\nu]],\qbrst),H_\nu
\end{eqnarray*}
is a contraction.
\end{theorem}
\begin{proof} 
Since the requisite condition $\res_\nu\;\qhop=0$ is obviously fulfilled, we apply 
Lemma \ref{BPL1} to the perturbation $\qbrst$ of $2\qkos$. Explicitly, this means that 
$H_\nu:=\frac{1} {2} \qhop\sum_{j=0}^\ell(-\frac{1}{2})^j(\qhop\qce+\qce \qhop)^j $ and 
$\Phi_\nu=\prol-H_\nu(\qce\;\prol-\prol\:\qdz)$, which are obviously series of Fr\'echet 
continuous maps. Note that from $h_0 \prol=0$ and $h^2=0$, we get $H_\nu\Phi_\nu=0$ and 
$H_\nu^2=0$. If $\prol$ is chosen to be equivariant, then the expression for $\Phi$ simplifies 
to $\Phi_\nu=\prol$. If $h$ and (hence $\qhop$) is equivariant, then it follows that 
$H_\nu=\frac{1}{2}\qhop$.
\end{proof}

We use this contraction to transfer the associative algebra structure from $\mathscr A[[\nu]]$ to the Lie algebra cohomology $\mathrm H^\bullet\big(\mathfrak g,\mathcal C^\infty(Z)[[\nu]]\big)$ of the representation $\mathbb L^z$
by setting 
\begin{eqnarray}\label{Formel1}
  [a]*[b]:=[\res_\nu\big(\Phi_\nu(a)*\Phi_\nu(b)\big)]
\end{eqnarray} 
where $[a],[b]$ denote the cohomology classes of 
$a,b\in C^\bullet\big(\mathfrak g, \mathcal C^{\infty}(Z)[[\nu]]\big)$. But in fact that is 
\emph{not exactly}, what we want to accomplish.
The primary obstacle on the way to the main result, Corollary \ref{redprod}, is that, in general, 
we have $\mathrm H^\bullet\big(\mathfrak g,\mathcal C^\infty(Z)[[\nu]]\big)\neq \mathrm H^\bullet\big(\mathfrak g,\mathcal C^\infty(Z)\big)[[\nu]]$. An example where this phenomenon occurs was 
given in \cite[section 7]{BHW}. One way out is to sharpen the compatibility condition (\ref{qmomentmap}).  
We require, that $\mathbb J=J$ and
\[J(X)\star f- f \star J(X)=\nu\{J(X),f\}\quad\mbox{ for all }X\in \mathfrak g, 
  f\in \mathcal C^{\infty}(M).\]
This property is also referred to as \emph{strong invariance} of the star product $\star$ with 
respect to the Lie algebra action. For proper group actions a strongly invariant star product 
can always be found (see \cite{Fed}). Of course, now the representations $\mathbb L$ and 
$\mathrm L$ coincide and we get $\delta=\qce$. But with some mild restrictions on the 
contracting homotopy $h$ of the Koszul resolution we also have the following.

\begin{lemma} 
  If $h_0$ is $\mathfrak g$-equivariant and $h_0\prol=0$, then 
  $\mathbb L^z=\mathrm L^z$.
\end{lemma}
\begin{proof} For $X\in\mathfrak g$ we have 
  $\mathbb L^z_X=\res_\nu\;\mathrm L_X \;\prol=\res\;(\id+(\qkos_1 -\partial_1)h_0)^{-1}
  \mathrm L_X\;\prol$. Since $\mathrm L_X$ commutes with $\qkos_1$, $\partial_1$ and $h_0$ the 
  last expression can be written as 
  $\res\;\mathrm L_X(\id+(\qkos_1 -\partial_1)h_0)^{-1}\prol=\res\;\mathrm L_X \;\prol$. 
\end{proof}

\begin{corollary} 
  \label{redprod} With the assumptions made above the product defined by equation 
  (\ref{Formel1}) makes $\mathrm H^\bullet\big(\mathfrak g,\mathcal C^\infty(Z)\big)[[\nu]]$ into a graded associative algebra. For the subalgebra  
$\mathrm H^{0}\big(\mathfrak g, \mathcal C^{\infty}(Z)\big)[[\nu]]= \big(\mathcal C^{\infty}(Z)\big)^{\mathfrak g}[[\nu]]$ this formula simplifies to 
\begin{eqnarray}\label{Formel2}
f*g:=\res_\nu\big(\prol(f)*\prol(g)\big)\quad\mbox{ for }f,g\in\big(\mathcal C^{\infty}(Z)\big)^{\mathfrak g}.
\end{eqnarray}
Since $\big(\mathcal C^{\infty}(Z)\big)^{\mathfrak g}[[\nu]]$ is
$\mathbb K[[\nu]]$-linearly isomorphic to the algebra of smooth functions 
on the symplectic stratified space $M_{\mathsf{red}}$, we obtain an associative
product on $\mathcal C^\infty(M_{\mathsf{red}})[[\nu]]$ which gives rise to a \emph{continuous} Hochschild cochain. 
\end{corollary}
\begin{proof} It remains to show (\ref{Formel2}). Let us denote by $\mathscr A^+$ the kernel of the augmentation map $\mathscr A\to \mathcal C^\infty(M)$. Equation (\ref{Formel2}) follows from the fact that $(\mathscr A^+\cap\mathscr A^0)[[\nu]]$ is a two-sided ideal in $\mathscr A^0[[\nu]]$.
\end{proof}
Finally, if $\mathrm H^1(\mathfrak g, C^\infty(Z))$ vanish, it is possible to find a topologically linear isomorphism between the spaces of invariants for the classical and the deformed representation. 
\begin{corollary} Let $G$ be a compact, connected semisimple Lie group acting on the Poisson manifold $M$ in a Hamiltonian fashion. Assume that the equivariant moment map $J$ satisfies the generating and the complete intersection hypothesis. Then for a star product $*$ on $M$ with quantum moment map $\mathbb J$ there is an invertible sequence of continuous maps \[S=\sum_{i\ge 0}\nu^i\:S_i :\mathrm H^0(\mathfrak g, \mathcal C^\infty(Z))[[\nu]]=\mathcal C^\infty(Z)^\mathfrak g[[\nu]]\to \mathrm H^0(\mathfrak g,\mathcal C^\infty(Z)[[\nu]])\] such that the formula 
\begin{align*}
f\star g:= S^{-1}\big(S(f)*S(g)\big)=S^{-1}\Big(\res_\nu\big(\Phi_\nu(S(f))*\Phi_\nu(S(g))\big)\Big)   
\end{align*}
defines a continuous formal deformation of the Poisson algebra $\mathcal C^\infty(Z)^\mathfrak g$ into an associative algebra. 
\end{corollary} 
\begin{proof} According to Viktor L. Ginzburg (see \cite[Theorem 2.13]{Ginz98}) we have for any compact, connected Lie group $G$ with a smooth representation on a Fr\'echet space $W$ an isomorphism 
$\mathrm H^\bullet(\mathfrak g,W)\cong \mathrm H^\bullet(\mathfrak g,\mathbb K)\otimes W^\mathfrak g$.
In particular, this implies that if $\mathfrak g$ is semisimple, the first and the second cohomology groups of the $\mathfrak g$-module $\mathcal C^\infty(Z)$ vanish. Note, that, since $G$ is compact, the space of invariants $\mathcal C^\infty(Z)^\mathfrak g\subset \mathcal C^\infty(Z)$ has a closed complement. This can be taken to be the kernel of the averaging projection. Using these observations it is straight forward  to construct $S$ by a standard inductive argument (see e.g. \cite[p.140]{BHW}). 
\end{proof}
\begin{appendix}
\section{Two perturbation lemmata} \label{HPT}
We consider (cochain) complexes in an additive $\mathbb K$-linear category $\mathscr C$ 
(e.g. the category of Fr\'echet spaces). A \emph{contraction} in $\mathscr C$ consists of the following data
\begin{eqnarray}\label{contr}
(X,d_X)\:
\begin{array}{c} {p}\\\leftrightarrows\\ {i}
\end{array}
\:(Y,d_Y),h_Y,
\end{eqnarray}
where $i$ and $p$ are chain maps between the chain complexes $(X,d_X)$ and $(Y,d_Y)$, $h_Y:Y\to Y[-1]$ is a 
morphism, and we have
$pi=\id_X$, $d_Yh_Y+h_Yd_Y=\id_Y-ip$. The contraction is said to satisfy the \emph{side conditions} (sc1--3), 
if moreover, $h_Y^2=0$, $h_Yi=0$ and $ph_Y=0$ are true. It was observed in \cite{LambeSt}, that in order to 
fulfill (sc2) and (sc3), one can replace $h_Y$ by $h'_Y:=(d_Yh_Y+h_Yd_Y) \:h_Y\:(d_Y h_Y+h_Y d_Y)$. 
If one wants to have in addition (sc1) to be satisfied, one may relapce $h_Y'$ by $h_Y'':=h_Y'd_Yh'_Y$. 
Let $C:=\operatorname{Cone}(p)$ be the mapping cone of $p$, i.e. $C=X[1]\oplus Y$ is the complex with 
differential $d_C(x,y):=(d_Xx+(-1)^{|y|}py,d_Y y)$. The homology of  $C$ is trivial, because 
$h_C(x,y):=(0,h_Yy+(-1)^{|x|}ix)$ is a contracting homotopy, i.e.~$d_Ch_C+h_Cd_C=\id_C$, if (sc3) is true.

Let us now assume that the objects $X$ and $Y$ carry complete descending filtrations and the structure maps are filtration preserving. Moreover, pretend that we have found a \emph{perturbation} $D_Y=d_Y+t_Y$ of $d_Y$, i.e. $D_Y^2=0$ and $t_Y:Y\to Y[1]$, called the \emph{initiator}, has the property that $t_Yh_Y+h_Yt_Y$ raises the filtration. Since, in general, $t_X:=pt_Yi$ needs not to be a perturbation of $d_X$, we impose that as an extra condition: we \emph{assume} that $D_X=d_X+t_X$ is a differential. Setting $t_C:=(t_X,t_Y)$, we will get a perturbation $D_C:=d_C+t_C$ of $d_C$, if we have  in addition $t_Xp=pt_Y$ (this will imply that $(d_X+t_X)^2=0$). Then an easy calculation yields that $H_C:=h_C(D_Ch_C+h_CD_C)^{-1}=h_C(\id_C+t_Ch_C+h_Ct_C)^{-1}$ is well defined and satisfies $D_CH_C+H_CD_C=\id_C$. Defining the morphism $I:X\to Y$, $H_C(x,0)=:(0,(-1)^{|x|}Ix)$ and the homotopy $H_Y:Y\to Y[-1]$, $H_C(0,y)=:(0,H_Yy)$ we get the following 
\begin{lemma}[\emph{Perturbation Lemma -- Version 1}] \label{BPL1}If the contraction (\ref{contr}) satisfies (sc3) and  $D_Y=d_Y+t_Y$ is a perturbation of $d_Y$ such that $t_Xp=pt_Y$, then 
\begin{eqnarray}\label{contrEins}
(X,D_X)\:
\begin{array}{c} {p}\\\leftrightarrows\\ {I}
\end{array}
\:(Y,D_Y),H_Y,
\end{eqnarray}
is a contraction fulfilling (sc3). Moreover, we have $H_Y= h_Y(\id_Y+t_Yh_Y+h_Yt_Y)^{-1}$ and $Ix=ix-H_Y(t_Yix-it_Xx)$. If all side conditions are true for (\ref{contr}), then they are for (\ref{contrEins}), too.
\end{lemma}

Starting with the mapping cone $K=\operatorname{Cone}(i)$, i.e. the complex $K=Y[1]\oplus X$ with the 
differential $d_K(y,x)=(d_Yy+(-1)^{|x|}ix,d_Xx)$, we may give a version of the above argument arriving at a 
contraction with all data perturbed except $i$. More precisely, we have a homotopy 
$h_K(y,x):=(h_Yy,(-1)^{|y|}py)$, for which $d_Kh_K+h_Kd_K=\id_K$ follows from (sc2). 
Mimicking the above argument, we get a differential $D_K:=d_K+t_K$ with $t_K:=(t_Y,t_X)$, if $t_Yi=it_X$ 
(this will imply $D_X^2=0$). Assuming (\ref{contr}) to satisfy (sc2), $H_K:=h_K(D_Kh_K+h_KD_K)^{-1}$ will 
become a contracting homotopy $D_KH_K+H_KD_K=\id_K$. Defining $P:Y\to X$ and $H'_Y:Y\to Y[-1]$ by  
$H_K(y,0)=H_K(y,x)=:(H'_Yy,(-1)^{|y|}Py)$ we get the following 
\begin{lemma} [\emph{Perturbation Lemma -- Version 2}] \label{BPL2}If the contraction (\ref{contr}) satisfies (sc2) and $D_Y=d_Y+t_Y$ is a perturbation of $d_Y$ such that $t_Yi=it_X$, then 
\begin{eqnarray}\label{contrZwei}
(X,D_X)\:
\begin{array}{c} {P}\\\leftrightarrows\\ {i}
\end{array}
\:(Y,D_Y),H'_Y,
\end{eqnarray}
is a contraction fulfilling (sc2). Moreover, we have $H'_Y= h_Y(\id_Y+t_Yh_Y+h_Yt_Y)^{-1}$ and $P=p(\id+t_Yh_Y+h_Yt_Y)^{-1}$. If all side conditions are true for (\ref{contr}), then they are for (\ref{contrZwei}), too.
\end{lemma}

\end{appendix}
\nocite{BHW,SjLerm,LerMontSj,AGJ}
\bibliographystyle{amsplain}
\bibliography{stratbrst}
\end{document}